\pdfoutput = 1
\documentclass[
    reprint, 
    amsmath,
    amssymb,
    aps,
    prb,
    superscriptaddress,
    showkeys,
    floatfix
]{revtex4-2}

\usepackage{graphicx} 
\usepackage{dcolumn}% Align table columns on decimal point
\usepackage{bm}
\usepackage{multirow}
\usepackage{float}
\usepackage{textgreek}
\usepackage[version=3]{mhchem}
\usepackage{xcolor}
\usepackage{hyperref}
\usepackage[mathlines, pagewise]{lineno}
\usepackage[american]{babel}
\usepackage{url}
\usepackage{soul}

\definecolor{JL}{rgb}{0.0,0.0,0.0}\def\JL{\color{JL}}

\definecolor{red}{rgb}{0.6,0.1,0.1}\def\red{\color{red}}

\begin{document}

\title{
Generalized Algorithm for Recognition of Complex Point Defects in Large-Scale \texorpdfstring{\textit{\textbeta}-\ce{Ga2O3}}{}
}

\author{Mengzhi Yan}
\affiliation{State Key Laboratory of Precision Measuring Technology $\&$ Instruments and Laboratory of Micro/Nano Manufacturing Technology, Tianjin University, Tianjin 300072, China}

\author{Junlei Zhao} 
\email{zhaojl@sustech.edu.cn}
\affiliation{Department of Electrical and Electronic Engineering, Southern University of Science and Technology, Shenzhen 518055, China}

\author{Flyura Djurabekova} 
\affiliation{Department of Physics and Helsinki Institute of Physics, University of Helsinki, P.O. Box 43, FI-00014, Finland}

\author{Zongwei Xu}
\email{zongweixu@tju.edu.cn}
\affiliation{State Key Laboratory of Precision Measuring Technology $\&$ Instruments and Laboratory of Micro/Nano Manufacturing Technology, Tianjin University, Tianjin 300072, China}
 
\date{\today}

\begin{abstract}

{\JL The electrical and optical properties of semiconductor materials are profoundly influenced by the atomic configurations and concentrations of intrinsic defects.
This influence is particularly significant in the case of $\beta$-\ce{Ga2O3}, a vital ultrawide bandgap semiconductor characterized by highly complex intrinsic defect configurations. 
Despite its importance, there is a notable absence of an accurate method to recognize these defects in large-scale atomistic computational modeling. 
In this work, we present an effective algorithm designed explicitly for identifying various intrinsic point defects in the $\beta$-\ce{Ga2O3} lattice. 
By integrating particle swarm optimization and hierarchical clustering methods, our algorithm attains a recognition accuracy exceeding 95\% for discrete point defect configurations. 
Furthermore, we have developed an efficient technique for randomly generating diverse intrinsic defects in large-scale $\beta$-\ce{Ga2O3} systems. 
This approach facilitates the construction of an extensive atomic database, crucially instrumental in validating the recognition algorithm through a substantial number of statistical analyses. 
Finally, the recognition algorithm is applied to a molecular dynamics simulation, accurately describing the evolution of the point defects during high-temperature annealing. 
Our work provides a useful tool for investigating the complex dynamical evolution of intrinsic point defects in $\beta$-\ce{Ga2O3}, and moreover, holds promise for understanding similar material systems, such as \ce{Al2O3}, \ce{In2O3}, and \ce{Sb2O3}.} 

\end{abstract}

\maketitle

\section{Introduction} \label{sec:intro}

{\JL $\beta$-Gallium oxide ($\beta$-\ce{Ga2O3}) has recently emerged as a vital candidate of ultrawide bandgap semiconductors.
Its distinct features, including a ultrawide bandgap of $4.8-4.9$ eV~\cite{2018_Ga2O3review}, a high and tunable $n$-type conductivity~\cite{varley2010oxygen, 2023_sidoping}, and the wide availability of high-quality bulk~\cite{heinselman2022projected, galazka2022growth} and thin-film~\cite{zavabeti2017liquid, vogt2017matal, itoh2020epitaxial, meng2022high, macco2022atomic} growth methods, underscore its potential applications in solar-blind ultraviolet optoelectronics~\cite{kim2020highly, hou2022high, zhang2023over} and high-voltage power electronics~\cite{zhang2022ultra, he2022over, zhou2023avalanche}.}

\begin{figure}[ht!] 
    \includegraphics[width=8.6cm]{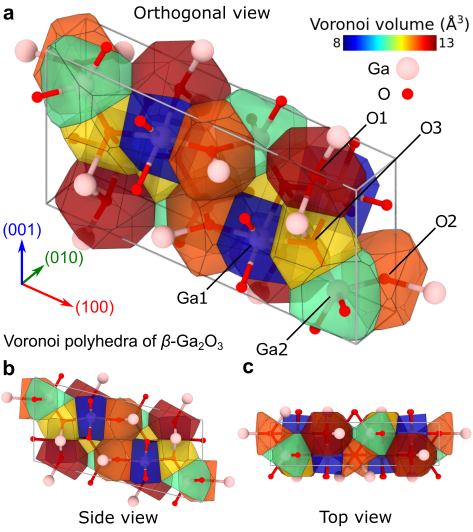}
    \caption{
    A 20-atom $\beta$-\ce{Ga2O3} conventional cell mapped with Voronoi polyhedral. 
    The Ga and O atoms are in pink and red, respectively. 
    The color coding of the polyhedral indicate their volumes. 
    The significant anisotropy of $\beta$-\ce{Ga2O3} lattice leads to the pronounced differences in the Voronoi polyhedral of various atoms, therefore, the commonly used WS point defect analysis method become inaccurate for analyzing $\beta$-\ce{Ga2O3} defects.
    Specific test results can be detailed in {\red Supplemental Material (SM) Appendix A}.
    }
    \label{fig:Voronoi_polyhedra}
\end{figure}

{\JL However, in contrast to other conventional semiconductors such as \ce{Si}, \ce{GaN}, \ce{SiC}, and diamond, the low-symmetry monoclinic lattice structure of $\beta$-\ce{Ga2O3} ($C2/m$, space group 12) poses an emerging challenge.
As illustrated in Fig.~\ref{fig:Voronoi_polyhedra}, a 20-atom conventional cell of $\beta$-\ce{Ga2O3} comprises (i) three types of \ce{O} sites, with O1 and O2 being 3-coordinated and the O3 being 4-coordinated; and (ii) two types of Ga sites, where Ga1 is 6-coordinated and Ga2 is 4-coordinated.
These intricate local atomic sties give rise to a widely diverse array of intrinsic point defect configurations, including simple \ce{Ga}/\ce{O} vacancies, split (or three-split) \ce{Ga} vacancies, 19 types of \ce{Ga}-\ce{O} divacancies, and regular/split \ce{Ga} interstitials~\cite{varley2010oxygen, 2019_Pointdetect, 2020_SplitGa, 2021_divacancies, 2023_Ga2O3_defects}.
Such intrinsic point defects can significantly impact the electrical and optical properties of $\beta$-\ce{Ga2O3}-based devices by acting as deep donors (\textit{e.g.}, \ce{O} vacancies, $\mathrm{V}_\mathrm{O}$~\cite{2018_VO_dft, 2022_O_formation_energy}) or shallow acceptors (\textit{e.g.}, \ce{Ga} interstitials, $\mathrm{Ga}_\mathrm{i}$~\cite{ingebrigtsen2019impact, 2020_Ga2O3_intrinsic_d, 2023_Ga2O3_defects}).
Therefore, a in-depth understanding and precise engineering of these intrinsic point defects in a large-scale dynamical system are crucial for the \ce{Ga2O3}-based applications.}

{\JL For large-scale atomistic computational modelling of solid lattice system, the Wigner-Seitz (WS) defect analysis method is conventionally employed to identify intrinsic point defects~\cite{wignerseitz1933, hammond2020parallel}.
The WS method relies on constructing referencing Voronoi polyhedra, which are spaces surrounded by perpendicular bisecting planes for all adjacent atoms in the reference configuration.
This approach is effective and computationally efficient for high-symmetry, isotropic lattices such as face-centred cubic, body-centred cubic, hexagonal close-packed, diamond, and various hexagonal stacking (\textit{e.g.}, $4H$ and $6H$) systems. 
However, as illustrated in Fig.~\ref{fig:Voronoi_polyhedra}, the low-symmetry, anisotropic $\beta$-\ce{Ga2O3} lattice results in a large diversity of the volume and shape of the Voronoi polyhedral. 
Moreover, some abundant and vital point detect types, such as split Ga vacancies and interstitials, cannot be accurately distinguished by the WS method.  
Hence, there is a pressing need to develop an efficient and reliable algorithm capable of recognizing complex point defects in $\beta$-\ce{Ga2O3}, and suited for the large-scale (\textit{e.g.}, $10^{3}-10^{6}$ atoms) dynamic modelling, such as molecular dynamics (MD) and kinetic Monte Carlo.}

{\JL In this contribution, we employ an analogous radial distribution function (ARDF) to identifying the local atomic environment.
For model refinement, we utilize particle swarm optimization (PSO) to enhance distinctions of standard configurations.
Subsequently, the unsupervised learning method of hierarchical clustering (HC) is applied in the secondary screening results.
The algorithm is validated through testing with a substantial number of static cells containing diverse \ce{Ga} point defect configurations. 
Finally, we explore the reliability and utility of the algorithm when deployed to monitor the defect evolution in a fully dynamic high-temperature annealing MD simulation.} 

\begin{figure*}[htbp]
    \includegraphics[width=16cm]{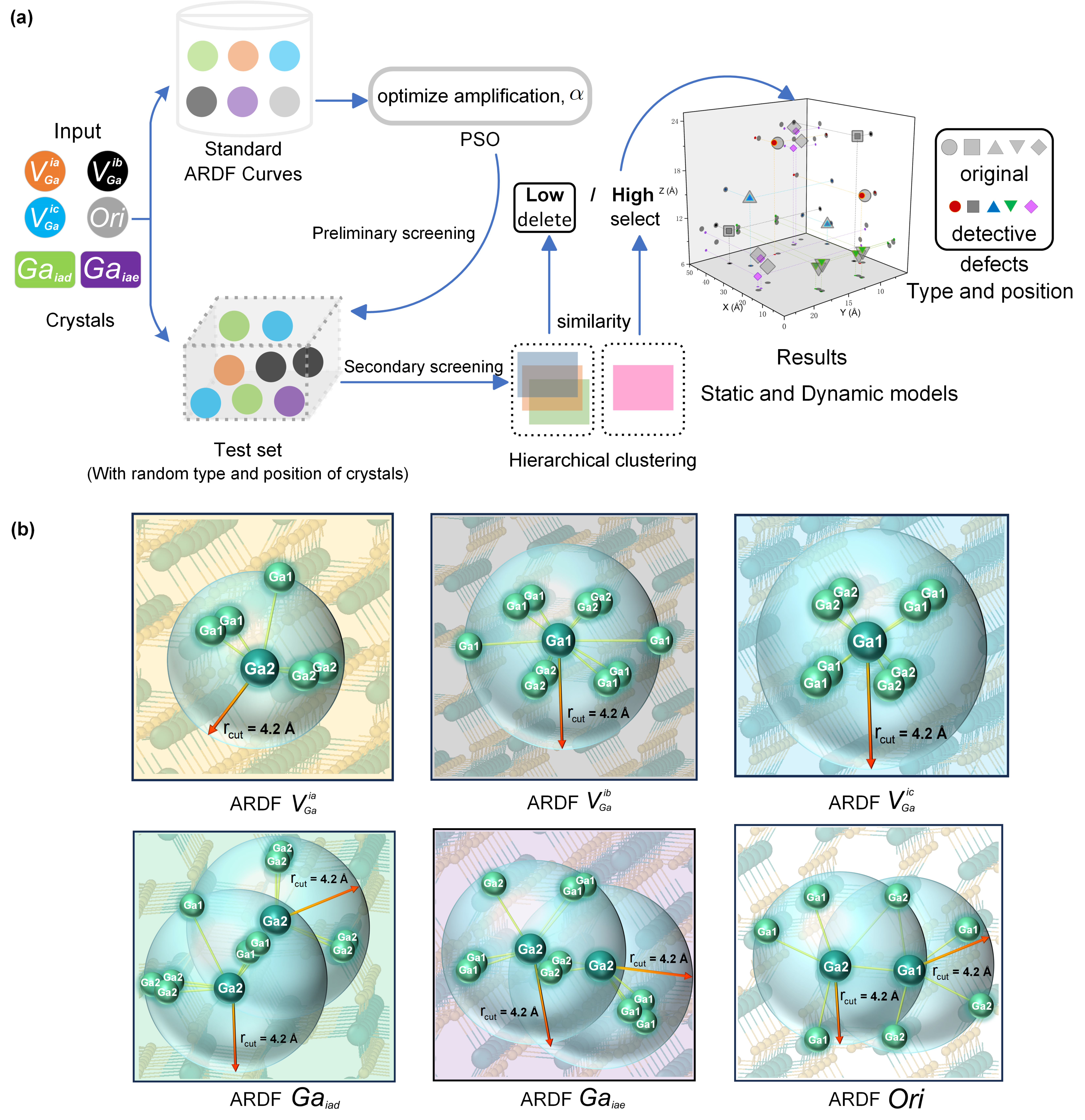}
    \caption{The flow process of the algorithm and the standard RDF curve.
    (a) A schematic diagram of the overall process of the identification algorithm. Arrow shows the information flow between the various components. 
    During preliminary screening, the basic structure of the defect configurations are constructed and used for the calculation of the standard pair radial distribution function (PRDF) curves. 
    The particle swarm optimization (PSO) method is then employed to determine the amplification coefficient $\alpha$, ensuring the maximum group distance between the standard data. 
    Various test sets are created, and the initial magnification is obtained after the relaxation process. 
    In secondary screening, hierarchical clustering (HC) method is applied for classifying results from the first step to obtain a cluster group with high similarity. 
    The final outputs are the point defect types and positions.
    (b) Local atomic environments for constructing standard ARDF curves: in the background, green and yellow atoms represent the Ga and O atoms, respectively. 
    A pale-colored shell denotes the designed maximum cutoff radius of 4.2 \r A. 
    The dark green highlighted atoms represent all the Ga atoms contained within the cutoff radius of the corresponding defect configurations.
    $\mathrm{Ga}1$ designates the 6-coordination Ga atom , while $\mathrm{Ga}2$ designates the 4-coordination Ga atom.
}
    \label{fig:flow_chart}
\end{figure*}

\section{Methodology} \label{sec:method}

\subsection{Dataset of Ga point defects} \label{subsec:dataset}

As summarized in Fig.\ref{fig:flow_chart}a, firstly, we construct reference atomic configurations of three split \ce{Ga} vacancies ($\mathrm{V}_\mathrm{Ga}^{i}$) and two \ce{Ga} interstitials ($\mathrm{Ga}_{i}$) with low formation energies~\cite{2019_Pointdetect, 2022_O_formation_energy, 2023_Ga2O3_defects}. 
the nonequivalent sites are labeled as difference types, namely, $\mathrm{V}_\mathrm{Ga}^{ia}$,$\mathrm{V}_\mathrm{Ga}^{ib}$, $\mathrm{V}_\mathrm{Ga}^{ic}$, $\mathrm{Ga}_{iad}$, and $\mathrm{Ga}_{iae}$, adopted from the notation in Ref.~\cite{2023_Ga2O3_defects}. 
The input data ``Ori" in Fig.~\ref{fig:flow_chart} refer to the atomic configurations of the two perfect \ce{Ga} sites.
{\JL We employ the ARDF, $g_\mathrm{A}(r)$, to describe the local atomic environment of the centered Ga atoms within a sphere of cutoff radius, $r$, as defined as follows: 
\begin{equation} \label{eq:ARDF}
    g_\mathrm{A}(r)= \frac{N_\mathrm{Ga}(r)}{V(r)},
\end{equation} 
where $N_\mathrm{Ga}(r)$ is the number of neighbouring \ce{Ga} atoms inside the sphere, and $V(r)=(4/3)\pi r^{3}$ is the volume of the sphere, as illustrated in detail in Fig.~\ref{fig:flow_chart}b.}
Notably, these reference ARDF curves can be constructed based on either \textit{ab initio} calculation~\cite{2023_Ga2O3_defects} or machine-learned classical method~\cite{2023_energy_4_founda}. 
Different methods yield marginal differences in the \ce{Ga} defect configurations, and hence, in the corresponding ARDFs. 
Nevertheless, the sensitivity and accuracy of our recognition algorithm, as elucidated in the following sections, are independent of these differences.
For consistency, in this work, we use the tabulated Gaussian approximation potential (tabGAP) from Ref.~\cite{2023_energy_4_founda} run with LAMMPS package~\cite{1995_lammps} to construct the input and test datasets.

\subsection{Recognition algorithm for Ga point defects} \label{subsec:algo}

The overall working principle of our recognition algorithm is to quantify the degree of similarity between any arbitrary ARDFs of initially unknown \ce{Ga} atoms and the standard ARDF curves, with high sensitivity and accuracy.
As such, these unknown \ce{Ga} atoms can be categorized into the known defect types or perfect sites.  
When comparing the ARDFs of the unknown \ce{Ga} atoms with the standard ARDFs, the discrete difference between the two curves, $\langle\overline{d}\rangle_{(a,b)}$, are defined as follows:
\begin{equation} \label{eq:access_similarity_1}
    \langle\overline{d}\rangle_{(a,b)} = \frac{1}{N} \sqrt{\sum_{n=1}^{N}\left[g_\mathrm{A,a}(r_{n}) - g_\mathrm{A,b}(r_{n})\right]^2},
\end{equation}
where $g_\mathrm{A,a}(r_{n})$ and $g_\mathrm{A,b}(r_{n})$ represent the ARDFs of the two \ce{Ga} atoms at shell radius of $r_{n} = (n/N)r_\mathrm{cut}$, and $N$ represents the total number of the discrete shells.
In this work, a cutoff radius, $r_\mathrm{cut}$, is set at 4.2 \r A (Fig.~\ref{fig:flow_chart}b) and a shell number, $N$, at 400. 
In this way, the similarity score, $S_{(a,b)}$, between the two ARDF curves $g_\mathrm{A,a}$ and $g_\mathrm{A,b}$ is defined as:
\begin{equation} \label{eq:access_similarity}
    S_{(a,b)} = \frac{1}{1+\alpha\cdot\langle\overline{d}\rangle_{(a,b)}},
\end{equation}
where $\alpha$ is amplification coefficient that determine the weight of $\langle\overline{d}\rangle_{(a,b)}$. 
By adjusting the value of $\alpha$, the $S_{(a,b)}$ between the two curves can be tuned.
{\JL Therefore, firstly, the optimized $\alpha$, denoted as $\alpha_\mathrm{best}$, should be set to maximize the overall dissimilarity by reaching the maximal $S_\mathrm{total}$, the sum of the absolute differences between each $S_{(a,b)}$ pair in the standard dataset, as follows:}
\begin{equation}
    S_\mathrm{total} = \frac{1}{2} \sum_{(a,b) \neq (c,d)} \vert S_{(a,b)} - S_{(c,d)} \vert,
\end{equation}
where a factor of $1/2$ is included to cancel the double counting of reversed pairs.
For this purpose, we employ PSO algorithm~\cite{1995_particle_first, 1998_particle_modified} with randomly distributed initial particle positions, $X_{i}^{0}$, and zero initial velocities, $V_{i}^{0}$.
The iterative velocity of the particle $i$ in the $t$-th iteration, $V_{i}^{t}$, is formulated as follows:
\begin{equation} \label{eq:distance_modify}
    \begin{aligned}
        V_{i}^{t} = & wV_{i}^{t-1} + c_{1}r_{1}(\mathrm{P(best)}_{i} - X_{i}^{t-1}) \\ & + c_{2}r_{2}(\mathrm{G(best)}^{t-1} - X_{i}^{t-1}),
    \end{aligned}
\end{equation}
where $w$ is inertia weight of the velocity from the previous iteration, $X_{i}^{t}$ is the position information of the particle $i$ in $t$-th iteration, $c_{1}$ and $c_{2}$ are two learning rates, $r_{1}$ and $r_{2}$ are two random factors in the range of $[0,1]$, $\mathrm{P(best)}_{i}$ represents the best particle position in the history of the particle $i$, and $\mathrm{G(best)}^{t-1}$ represents the best particle positions among all the particles closest to the optimal solution in the ($t-1$)-th iteration. The position of the particle $i$ in $t$-th iteration, $X_{i}^{t}$ can be updated as:
\begin{equation}
\label{eq:distance_modify_1}
        X_{i}^{t} = X_{i}^{t-1} + V_{i}^{t-1},
\end{equation}
where $X_{i}^{t-1}$ is the  position of the particle $i$ in the ($t-1$)-th iteration. 
We note that the overall sensitivity of recognition is fairly good when the $\alpha$ is within the optimized range (Fig.~\ref{fig:amplified_parameters}). 
Therefore, the optimization is halted when the number of iterations reaches the preset maximum or the change of the best position among the particles, $\mathrm{G(best)}^{t}$, falls below the convergence threshold.
Table~\ref{tab:particle_swarm} summarize the parameters of the PSO algorithm to optimize the $\alpha$. 
 
\begin{table}[htbp]
\caption{Detailed parameters of the PSO algorithm to optimize the amplification coefficient, $\alpha$.}
    \begin{ruledtabular}
    \begin{tabular}{c c}
    Parameters & Values\\
    \hline
    Particle number                 & 50    \\
    Particle dimension              & 1     \\
    Maximum number of iterations    & 250   \\
    Inertia weight, $w$             & 0.5   \\
    Learning factors, $c_{1}$ and $c_{2}$    & 0.2   \\
    Random factors, $r_{1}$ and $r_{2}$    & $[0,1]$ \\
    Lower limit of solution space   & 1     \\
    Upper limit of solution space   & 30    \\
    Convergence threshold of $\mathrm{G(best)}^{t}$  & 0.0001  \\
    \end{tabular}
    \end{ruledtabular}
    \label{tab:particle_swarm}
\end{table}

The optimized $\alpha_\mathrm{best}$ is subsequently utilized to compute the similarity between the unknown ($\mathrm{Un.}$) \ce{Ga} particles and all the standard ($\mathrm{Std.}$) ARDFs, represented as $S_\mathrm{(Un.,Std.)}$. 
Through this approach, the abundant, perfect \ce{Ga} atoms are effectively screened under the condition of maximal similarity to the standard Ga1 or Ga2 sites.
This step is referred to as the `preliminary screening' process in Fig.~\ref{fig:flow_chart}a.
Notably, this process significantly enhances the computational efficiency of our algorithm, by substantially reducing the number of atoms  processed during the secondary screening. 

The aim of the secondary screening in our algorithm is to further categorize defect types and pinpoint their positions with high accuracy.
For this purpose, we employ the HC method~\cite{1988_her, 2001_elbow}, an unsupervised algorithm designed to handle an unknown number of categories. 
Specifically, our recognition algorithm uses an array consisting of nine $S_\mathrm{(Un.,Std.)}$ of a defective \ce{Ga} atom as a grouped input for clustering analysis.

An elbow diagram is employed to determine the optimal number of clusters. 
The $y$-axis of this plot represents the in-cluster sum of squared errors, denoted as $\mathrm{SSE}$:
\begin{equation} \label{eq:Coefficient of polymerization_1}
\mathrm{SSE} = \sum_{k=1}^\mathrm{K} \sum_{S_\mathrm{(Un.,Std.)} \in C_{k}} \vert S_\mathrm{(Un.,Std.)}-\mu_{k}\vert^2,
\end{equation}
where $k$ is the cluster index ($k=1,2,\dots,\mathrm{K}$), $C_{k}$ is the clustered set with $n_{k}$ elements, and $\mu_{k}$ represents the numerical-average center of the cluster $C_{k}$. $\mu_{k}$ is calculated as follows:
\begin{equation} \label{eq:center_value}
\mu_{k} = \frac{1}{n_{k}}\sum_{S_\mathrm{(Un.,Std.)}\in C_{k}}S_\mathrm{(Un.,Std.)},
\end{equation}
where $\mu_{k}=S_\mathrm{(Un.,Std.)}$ for a single-element cluster.

Subsequently, the optimal number of clusters is determined based on the inflection point observed in the elbow diagram (Fig.~\ref{fig:cluster_example}). 
The number of clusters ($k=1,2,\dots,\mathrm{K}$) showing the most significant change in the degree of distortion is selected as the $k$-nearest neighbor cluster number. 
Eventually, an inertia, $I$, is introduced to calculate the difference between unknown particle curves ($g_\mathrm{A,Un.}(r_n)$) and the standard ARDF curve ($g_\mathrm{A,Std.}(r_n)$) of the defect structures with the highest similarity obtained from the preliminary screening.
The inertia, $I$, is defined as follows:
\begin{equation} \label{eq:center_value}
I = \sum_{n=1}^{N}g_\mathrm{A,Un.}(r_n)-g_\mathrm{A,Std.}(r_n),
\end{equation}
where $N$ represents the total number of the discrete shells.
Then we obtain the clustering results of particles with different similarity and $I$ under the optimal number of clusters.
Ultimately, selecting high similarity points in the clustering results for defect point type and position statistics.
This enables the accurate detection and monitoring of \ce{Ga} defect types and positions.

% $S_\mathrm{(a,b)}$ is the similarity of each particle as a defect configuration obtained by PSO, $a$ is one ARDF line of a particle in test set while b is standard ARDF lines used to be compared. $\mu_{k}$ corresponds to the cluster center under a specific cluster number $k$ and $C_{k}$ represents all ARDF lines of different test particles in $k$ category, $n$ is the number of lines in category $k$, as depicted in Eq.~\ref{eq:center_value}.
% Subsequently, we determine the number of clusters ($k=1,2,\dots,\mathrm{K}$) based on the inflection point on the elbow diagram and $\mathrm{K}$ represents the total number of particles after the initial screening.

% HC with KNN is performed based on the number of clusters, as shown in Eq.~\ref{eq:Coefficient of polymerization_1},
% and there is a high probability of obtaining the group with low SSE, 
% representing the correct category and the corresponding number of defects for the identification points. 
% Following the aforementioned experimental concepts and algorithm design, we have successfully developed a method capable of setting the concentration of defect configurations and identifying the defect structure and the number of points.

\subsection{Test procedure} \label{sec:level5}

Static and dynamic test cells are designed to verify the reliability of our recognition algorithm.
Setting of system size and defect number are set as shown in the Table~\ref{tab:blockset}.

\begin{table}[htbp]
    \caption{Parameter setting of different test set}
    \begin{ruledtabular}
    \begin{tabular}{c c c c c c c}
    {\JL Test sets} & {\JL Atoms number} & {$\mathrm{V}^{ia}_\mathrm{Ga}$} &{$\mathrm{V}^{ib}_\mathrm{Ga}$} &{$\mathrm{V}^{ic}_\mathrm{Ga}$} &{$\mathrm{Ga}_{iad}$} &{$\mathrm{Ga}_{iae}$}\\
    \hline
    % Work material           & Monoclinic $\beta$-\ce{Ga2O3} plane\\
    % Potential function      & tabgap or soapgap\\
    $\mathrm{a}$ & 3998 & 1 & 1 & 1 & 1 & 1 \\
    $\mathrm{b}$ & 3999 & 2 & 2 & 2 & 2 & 2 \\
    $\mathrm{c}$ & 6001 & 2 & 4 & 1 & 5 & 3 \\
    \end{tabular}
    \end{ruledtabular}
    \label{tab:blockset}
\end{table}

In static test, the atomic configurations of the test sets are first relaxed to the local potential energy minimum.
\begin{figure*}[htbp]
    \includegraphics[width=16cm]{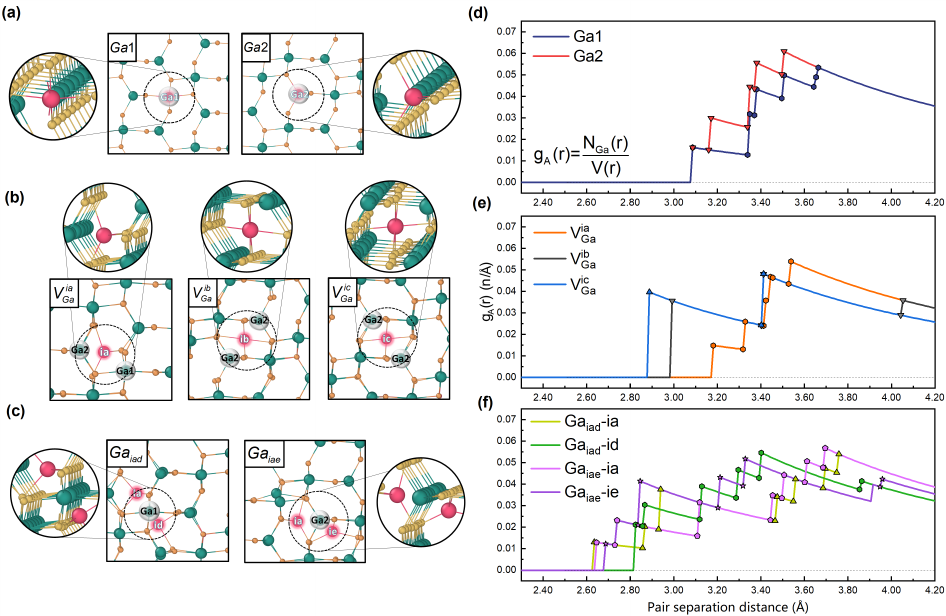}
    \caption{
    Atomic configurations of (a) perfect 6-coordination Ga1 atoms and 4-coordination Ga2 atoms, (b) three Ga vacancies, and (c) two split Ga interstitials. 
    (d, e, f) Their corresponding ARDFs. 
    The ARDF curves have a cutoff radius, $r_\mathrm{cut}$, of 4.2 \r A.}
    \label{fig:all_defects}
\end{figure*}
Utilizing the energy-stable frames from the dataset, and aiming for thermodynamically valid data, the average coordinates of each particle are calculated to serve as the raw data for the recognition object.

Similarly, to minimize interference from atomic lattice vibrations during the annealing process, 
data files are extracted at 900 K temperature and specified time steps. 
The same energy minimization process is applied to explore the evolution of the number and types of defect configurations during annealing. 
Then we detect the defect configuration of the stable process after energy minimization to obtain our final testing results.

% This leads to the loss of identification points, impacting experimental results. 
% To address this, we simplify the dynamic recognition process by selecting particle information at a specific annealing time. Using LAMMPS, 
% we minimize energy and denoise the system in a limited time step to accurately identify particles, including those at the boundary points.

% 第四部分

\section{RESULTS AND DISCUSSION} \label{sec:level6}

\subsection{Standard ARDF curves} 

Fig.~\ref{fig:all_defects} illustrates the standard ARDF curves corresponding to stable configurations of different perfect Ga sites and defects, labeled in accordance with previous works~\cite{2023_Ga2O3_defects,2019_Pointdetect}. 
In $\mathrm{V}_\mathrm{Ga}^{ia}$, $\mathrm{V}_\mathrm{Ga}^{ib}$, and $\mathrm{V}_\mathrm{Ga}^{ic}$ configurations, two Ga vacancies share a Ga atom, causing this Ga atom to be positioned between the vacancies, as shown in Fig.~\ref{fig:all_defects}b. 
Conversely, in $\mathrm{Ga}_{iad}$ and $\mathrm{Ga}_{iae}$ configurations, two Ga atoms share a Ga vacancy, as shown in Fig.~\ref{fig:all_defects}c. 
In Fig.~\ref{fig:all_defects}d-f, highlight differences among ARDF curves.
Differences are observed in ARDF curves within a 4.2 \r A ranging for Ga atoms with two distinct coordination numbers.
The maximum value of $g_\mathrm{A}(r)$ for Ga1 can reach 0.06, whereas for Ga2, the maximum value is only 0.05. 
Notably, a significant disparity exists in the curves for $\mathrm{V}_\mathrm{Ga}^{ia}$, $\mathrm{V}_\mathrm{Ga}^{ib}$, and $\mathrm{V}_\mathrm{Ga}^{ic}$ configurations of the split vacancy. 
The first Ga atom appears in the $\mathrm{V}_\mathrm{Ga}^{ic}$ configuration at approximately 2.9 \r A, in the $\mathrm{V}_\mathrm{Ga}^{ib}$ configuration at 3.0 \r A, and in the $\mathrm{V}_\mathrm{Ga}^{ia}$ configuration at about 3.2 \r A. 
Due to the local symmetry of the $\mathrm{V}_\mathrm{Ga}^{ic}$ configuration and the $\mathrm{V}_\mathrm{Ga}^{ib}$ configuration, it can be observed that their characteristic curves overlap in most cases. 
However, another Ga atom appears in the $\mathrm{V}_\mathrm{Ga}^{ib}$ configuration at around 4.05 \r A, resulting in an increased difference between it and the $\mathrm{V}_\mathrm{Ga}^{ic}$ configuration.
For the $\mathrm{V}_\mathrm{Ga}^{ia}$ configuration, the atomic environment of Ga distribution is significantly distinct from that of $\mathrm{V}_\mathrm{Ga}^{ib}$ and $\mathrm{V}_\mathrm{Ga}^{ic}$.
Additionally, since one split Ga interstitial corresponds to two defected Ga atoms, it is necessary to draw two ARDF curves for each interstitial to illustrate its features, as shown in Fig.~\ref{fig:all_defects}f.
A total of four ARDF curves are therefore presented for the $\mathrm{Ga}_{iad}$ and $\mathrm{Ga}_{iae}$ configurations.
\subsection{Amplification coefficient \texorpdfstring{\textit{\textalpha}}{}} 
\begin{figure*}[htbp] %% 这个图片要记得放出来
    \includegraphics[width=17cm]{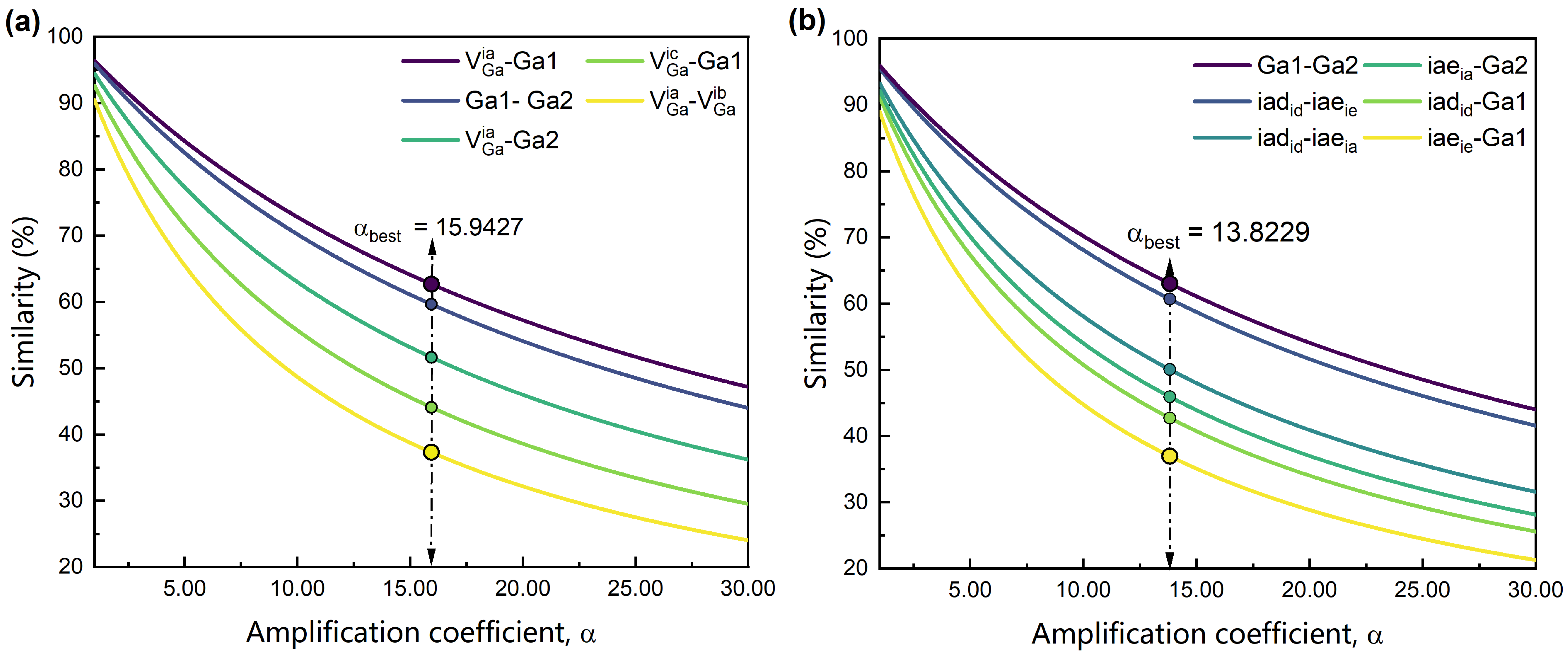}
    \caption{\label{fig:amplified_parameters}
    Optimization of amplification coefficient. (a) The function relation between the paired similarity, S, and the amplification coefficient, $\alpha$, among the three split Ga vacancy configurations and the referencing curves of the perfect 6-coordination ($\mathrm{Ga}1$) and 4-coordination ($\mathrm{Ga}2$) Ga sites. (b) The relation of the Ga split interstitials and the perfect Ga sites. The value of the optimized amplification coefficient, $\alpha_{best}$, is labelled by the dashed line where all the paired similarities size values are greatest.
    }
\end{figure*}
The differences
between standard databases obtained using the conventional euclidean distance
are very small.
In particular, for the split interstitial structure $\mathrm{Ga}_{iad}$ and $\mathrm{Ga}_{iae}$ configuration, the absolute difference between their curves is minimal, almost 0.05. 
This small difference could lead to a significant error in particle identification. 
To enhance accuracy and expand the distinction between the identification curves in the database, an amplification coefficient $\mathrm \alpha$ is introduced, as shown in Eq.~\ref{eq:access_similarity}.
By adjusting the value of $\mathrm \alpha$, the difference between the curves can be expanded.
In the case of the split vacancy, the database computes a total of 5 ARDF curves with perfect Ga1, Ga2 and 3 split vacancy structures, $\mathrm{V}_\mathrm{Ga}^{ia}$, $\mathrm{V}_\mathrm{Ga}^{ib}$ and $\mathrm{V}_\mathrm{Ga}^{ic}$. 

By calculating the similarity of paired curves among 5 different sets, a total of 10 sets of solutions are formed. 
We then calculate the amplification coefficient $\mathrm \alpha$ for the first set of 10 curves.
Regarding the split interstitial structures, 
configurations $\mathrm{Ga}_{iad}$ and $\mathrm{Ga}_{iae}$ have two distribution lines each, 
denoted as ia, id, and ia, ie, respectively. 
In addition to the Ga1 and Ga2 in the two perfect lattices, 
there are 6 ARDF curves, leading to 15 sets of results after paired combination, 
as mentioned earlier.

% 根据相似度计算方程可知，曲线在插值之后两两之间差异值是固定的，于是相似度和放大倍数alpha之间呈现倒数关系，可将对应结果绘制如图3所示，由于当α趋近于无穷时，曲线之差接近于0，故一定存在某一放大系数让曲线之间的差异值最大。计算中由于score1 - score2 + score 2 - score 3化简后即score1 - score 10 的最值
% 故我选择的目标函数便为score1 - score 10 的最大值即可，于是通过第二部分的粒子群算法的迭代之后，我们计算得到了ia，ib，ic和两种Ga配位结构，4配位和6配位之间的放大倍数为：15.0665；而另一组，计算了id1，id2和ie1，ie2曲线和两种完美晶格Ga的配位数之间的放大倍数为：13.6左右。

Through preliminary tests, it is observed that calculating the distance between two curves resulted in a maximum value. The sum and maximum of differences among the 10 groups of curves are not significantly different from directly calculating the difference between the top and bottom curves. 
The amplification coefficient $\mathrm \alpha$ obtained also showed minimal variation.
Consequently,
the sum of differences between two adjacent curves is approximated by calculating the difference between the top and bottom curves. 
According to the calculations, with the amplification coefficient ranging from 1 to 30, 
we compute the $\mathrm{V}_\mathrm{Ga}^{ia}$, $\mathrm{V}_\mathrm{Ga}^{ib}$, $\mathrm{V}_\mathrm{Ga}^{ic}$, and two Ga coordination structures. 
Fig.~\ref{fig:amplified_parameters}a illustrates that the amplification coefficient $\alpha_\mathrm{best}$ between the split vacancy structures and perfect Ga sites is 15.9427.
In parallel, the other group calculates the $\alpha_\mathrm{best}$ between ia, id, ia, ie curves, 
and the coordination number of the perfect Ga1 and Ga2 structures. 
Fig.~\ref{fig:amplified_parameters}b reveals that the magnification between the split interstitial structures and perfect Ga
sites is 13.8229.
The iteration flow of the PSO algorithm can be comprehended through {\red SM Appendix C}.
%%%%%%%%%%%%%%%%%%%%%%%%%%%%%%%%%%%%%%%%%%%%%%%%%%%%%%%%%%%%
% 在分别获得了分裂空位缺陷的识别阈值以及分裂间隙子的识别阈值后，将该值带入相似度函数计算测试对象中每一个粒子与数据库中标准缺陷构型的ARDF的相似度大小，将对应相似度最大的种类视为该粒子对应的构型情况。接着由于初次分类结果中，对于每一种类别相似度存在差异，并且根据实验观察，粒子与对应缺陷构型相似度大，说明该结果是我们所需要的缺陷构型，而相似度较小的情况所对应的是处于缺陷点位附近的4配位构型或者6配位构型。于是我们统计不同构型中所有粒子的相似度开始进行二次识别分类，根据相似度之间的大小关系，我们采用层次聚类的思想直接对余下的检测结果进行聚类，首先根据肘部图获得每一种构型所需的聚类数目，再采用该聚类数目与k-means方法，对大小不一的相似度进行有效的区分，
% 图4展示的是实验17提供的结果，可以发现我们的方法对于多种构型的肘部图结果可以提供每一种构型所需要的聚类数量信息，在对应的聚类数量下，绘制得到的聚类结果如右图所示，对于V_{ia}结构而言聚为两类分别如右图所示，相似度大小分别为：**，**。同理对于b构型的聚类结果为两类分别为：**，**。而c构型聚为一类其结果也如图所示，最后d和e构型则分为3类，d的聚类相似度分别为：***，***。e为***，***。
% 根据聚类结果可以证明，对于离散的Via，Vib，Vic，Gaiad以及Gaiae构型而言，识别准确率可以达到100%，而对于点缺陷构型较为集中的情况，该准确性有所下降，特别是对于构型Via而言。由于部分缺陷之间会相互影响，例如位于id点位的Ga原子和用于构成Via构型而位于ia点位的Ga原子，同时也会存在两种类型的缺陷构型相聚较近，导致二者形成了更加复杂的缺陷构型，尤其是分裂间隙子iad构型和分裂空位Via构型，它们会额外形成一个位于ib点位的构型，而导致原先二者的识别结果少于预期，但这种现象的发生也说明我们的算法并不会将相似的点位识别为对应缺陷的情况。
After obtaining the amplified coefficient $\alpha$ for split vacancy defects and split interstitial thresholds separately, these values are inserted into a similarity function to calculate the ARDF similarity magnitude between each particle in the test object and the standard defect configurations in the database. 

Subsequently, due to differences in similarity for each category in the initial classification results, and particles with high similarity to the corresponding defect configuration indicate that this result corresponds to the defect configuration. 
Simultaneously, there are particles identified as corresponding defect structures, but with low similarity.
As a result, Ga1 and Ga2 atoms in the perfect lattice are misidentified in the recognition results of split vacancy defect and split interstitial defect configurations.
Therefore, we perform a secondary screening of different configurations by calculating the similarities among all particles.
Based on the similarity, we employ a HC approach to directly cluster the remaining detection results. 

\begin{figure*}[htbp] % 这个图片要记得放出来
    \includegraphics[width=17cm]{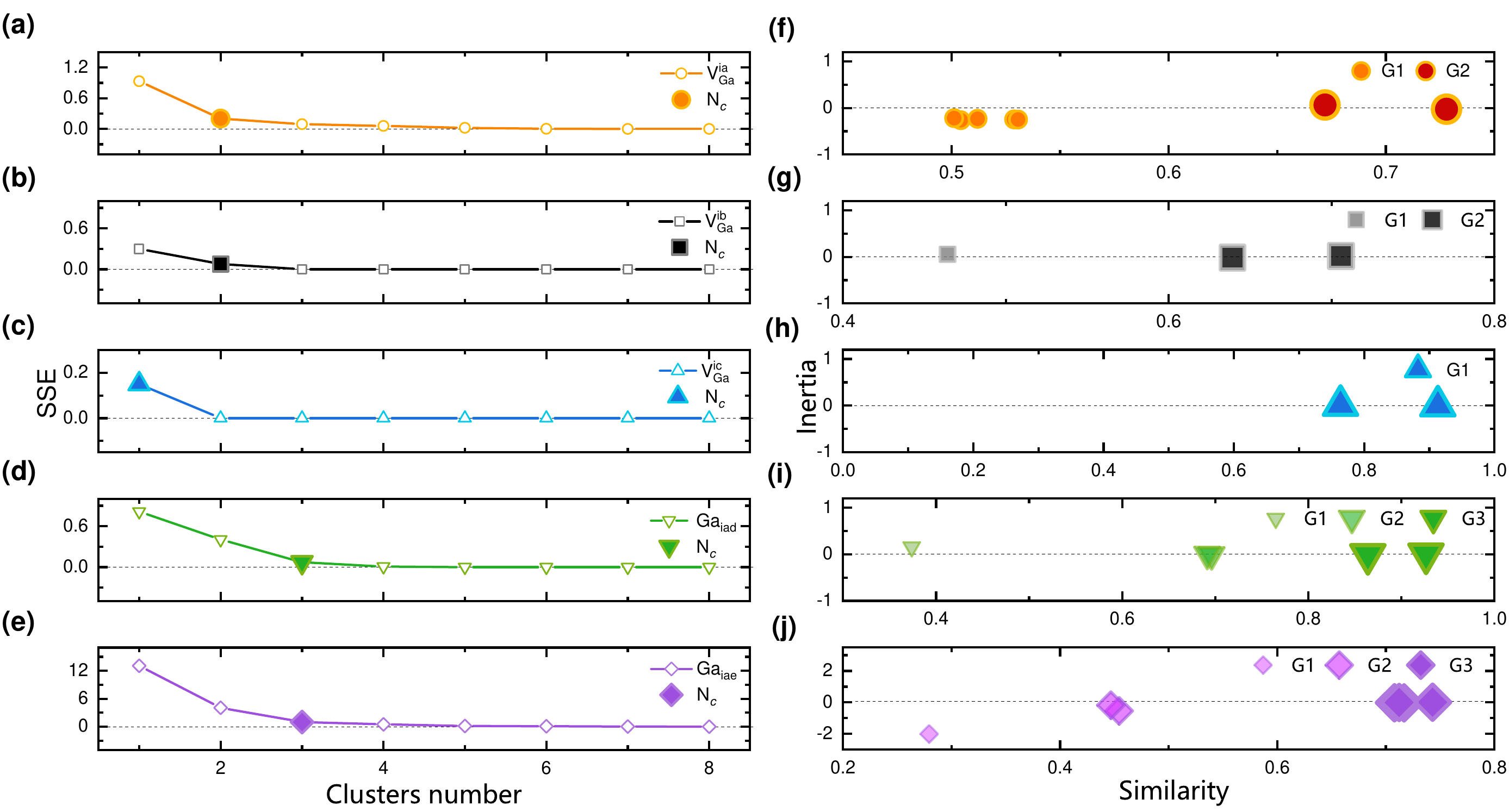}% Here is how to import EPS art
    \caption{\label{fig:cluster_example} HC results of the exemplary test. The left panels (a)-(e) show the Elbow diagrams for $\mathrm{V_\mathrm{Ga}^{ia}}$, $\mathrm{V_\mathrm{Ga}^{ib}}$, $\mathrm{V_\mathrm{Ga}^{ic}}$, $\mathrm{Ga_{iad}}$, and $\mathrm{Ga_{iae}}$ defect configurations, along with their respective clusters number selections. $\mathrm{N}_C$ denotes the choice of the cluster number. $\mathrm{SSE}$ denotes the sum of squared errors. The right panels (f)-(j) illustrate the clustering results obtained by  (a) - (e). G1, G2,\dots stand for group index. The clustering outcomes are categorized based on the degree of similarity.}
\end{figure*}

% 23.11.23 改写至此

\subsection{HC algorithm for clustering} 

Fig.~\ref{fig:cluster_example} shows the results provided by exemplary test set. 
HC method can provide the number of clusters required by each configuration for the elbow diagram results of various configurations~\cite{1998_particle_modified,1995_particle_first}. 

\begin{figure*}[htbp]
    \centering
    \includegraphics[width=17cm]{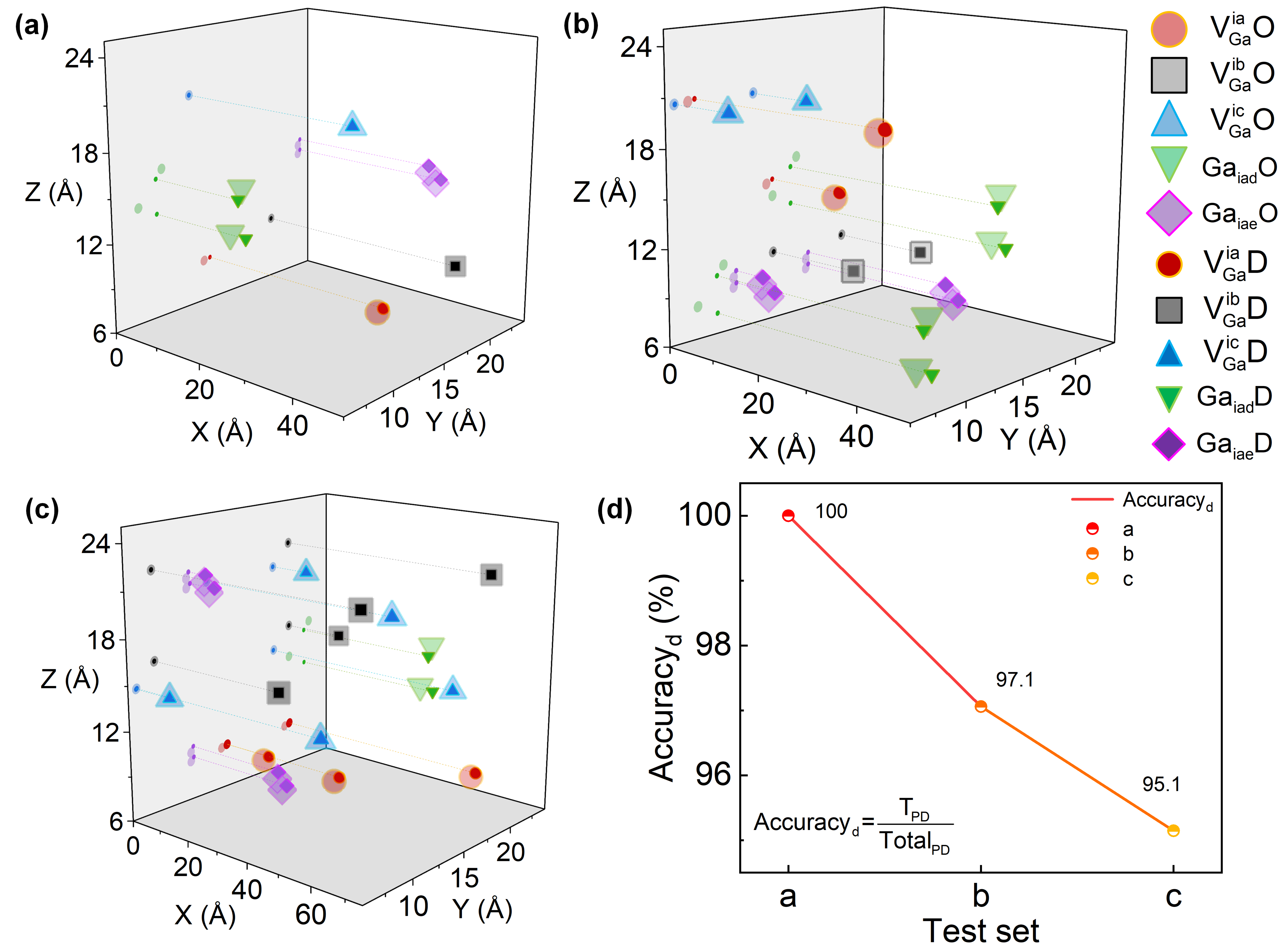}
    \caption{\label{fig:results}Exemplary test results of different random defective cells. (a) Five preset defect configurations (one for each type) in a 4000-atoms cell. (b) Ten defect configurations (two for each type) in a 4000-atoms cell. (c)  Fifteen defect configurations (random number for each type) in a 6000-atoms cell. (d) The accuracy of defect identification of the above three sets of test. `O' represents the original point location of the defect, and `D' represents the point location identified by the algorithm.}
\end{figure*}

Utilizing the relationship between the SSE and the number of clusters within different defect groups, as calculated by Eq.~\ref{eq:Coefficient of polymerization_1}, we plotted the SSE of different cluster numbers in Fig.~\ref{fig:cluster_example}a-e.
The clustering results are depicted in the right figure. Fig.~\ref{fig:cluster_example}f-j depicts the relationship between $I$ and the similarity obtained in the test set.
Considering the relative position of particles and the zero value in $I$, preliminary judgments can be made that the particles with $I$ close to zero are more similar to corresponding defects.

Specifically,
for $\mathrm{V}_\mathrm{Ga}^{ia}$ structures, 
the clustering is bifurcated into two categories, 
as shown in Fig.~\ref{fig:cluster_example}a and f
with similarity values of 0.515 and 0.700. 
Similarly, the clustering results for $\mathrm{V}_\mathrm{Ga}^{ib}$ configuration are divided into two categories, 
with similarity values of 0.464 and 0.672. 
The result of $\mathrm{V}_\mathrm{Ga}^{ic}$ configuration clustering, 
as shown in Fig.~\ref{fig:cluster_example}c and h, is 0.838. 
Finally, $\mathrm{Ga}_{iad}$ and $\mathrm{Ga}_{iae}$ configurations are segmented into three clusters, 
and the clustering similarities for $\mathrm{Ga}_{iad}$ are 0.374, 0.694, and 0.895, respectively, as shown in Fig.~\ref{fig:cluster_example}d and i.
In Fig.~\ref{fig:cluster_example}e and j, for $\mathrm{Ga}_{iae}$, the values are 0.279, 0.453, and 0.720.

% % 24.1.8 afternoon
% The $\mathrm{V}_\mathrm{Ga}^{ia}$ configuration, 
% due to its atomic environment closely resembling that of the 4-coordinated perfect lattice Ga, may cause algorithmic confusion. 
% To address this, the threshold for the $\mathrm{V}_\mathrm{Ga}^{ia}$ configuration is set to 0.9 times the maximum similarity, 
% designating any similarity below this value as indicative of a different structure. 
% Similarly, the $\mathrm{Ga}_{iad}$ configuration's lower recognition threshold is set to 0.7 times the maximum recognition result.

\subsection{Static and dynamic procedure} 
In the upcoming test, 
we delve into both the static and dynamic recognition processes of the algorithm. 
To ensure the randomness of the test set, the designed programme is utilized to splice and combine the initial defect configuration of about 80 particles with the perfect lattice structure.
Details of the input data can be found in {\red SM Appendix B}.
The concentration of different defect configurations, \textit{i.e.}, the number of defect input data varies in different test sets. 
In the defect detection of the static test set, 3 groups of tests are designed, as indicated in Table.~\ref{tab:blockset}. 
See {\red SM Appendix E} for more tests.
Averaging a stable number of steps for each test set provides the initial data for the test.

Fig.~\ref{fig:results} illustrates the perfect recognition results obtained by the algorithm for different total numbers of particles and various defect densities. 
Fig.~\ref{fig:results}a-c demonstrate recognition results under different conditions, 
showcasing the algorithm's ability to obtain accurate results for discrete and stable defect configurations. 
Notably, 
the completion of the algorithm design and conducting 40 sets of independent test, 
consistent recognition accuracy of 95\% or higher is observed for discrete point defect configurations $\mathrm{V}_\mathrm{Ga}^{ia}$, $\mathrm{V}_\mathrm{Ga}^{ib}$,  $\mathrm{V}_\mathrm{Ga}^{ic}$, $\mathrm{Ga}_{iad}$ and $\mathrm{Ga}_{iae}$.
These accuracy rates will be continually updated as test progress.

However, for configurations where point defects are more concentrated, this accuracy will slightly decreases.
In Fig.~\ref{fig:results}d, accuracy statistics for each group of test results is displayed. 
$\mathrm {T}_\mathrm{PD}$ denotes the number of defect points identified through algorithm feedback, 
while $\mathrm {Total}_\mathrm{PD}$ represents the overall number of defects introduced in the test.
Considering the clustering results, 
in a test set comprising a total of 4,000 particles, when point defects are relatively discrete, their count is only 5.
The recognition accuracy for  $\mathrm{V}_\mathrm{Ga}^{ia}$, $\mathrm{V}_\mathrm{Ga}^{ib}$, $\mathrm{V}_\mathrm{Ga}^{ic}$, $\mathrm{Ga}_{iad}$, and $\mathrm{Ga}_{iae}$ can achieve 100\%. 
With an increased defect number of 10, the algorithm's accuracy drops to 97.1\%. Subsequently, as the number of particles rises to 6000 and the total defects increase to 15, the recognition accuracy rate becomes 95.1\%.
Nevertheless, as the defect density and total number of particles in the system increase, the overall accuracy decreases, yet it still remains at 95\% or above.

With an increase in defect density, overall recognition efficiency decreases due to the compound phenomenon between nearby defects.
Particularly for the $\mathrm{V}_\mathrm{Ga}^{ia}$ configuration. 
Some defects may exert mutual influence. 
For instance, Ga atoms situated at the $\mathrm{Ga}_{iad}$ site and Ga atoms involved in forming the $\mathrm{V}_\mathrm{Ga}^{ia}$ configuration at the ia site can be in close proximity. 
This proximity may lead to the creation of more complex defect configurations, 
especially in the case of the $\mathrm{Ga}_{iad}$ configuration for split interstitial and the $\mathrm{V}_\mathrm{Ga}^{ia}$ configuration for split vacancy. 
This interplay can result in an additional configuration at the $\mathrm{V}_\mathrm{Ga}^{ib}$, leading to fewer expected recognition results for these configurations.
Refer to {\red SM Appendix D} for details.
This phenomenon's occurrence further illustrates that our algorithm accurately distinguishes similar sites, 
preventing misidentification as corresponding defects.

\begin{figure*}[htbp]
    \includegraphics[width=18cm]{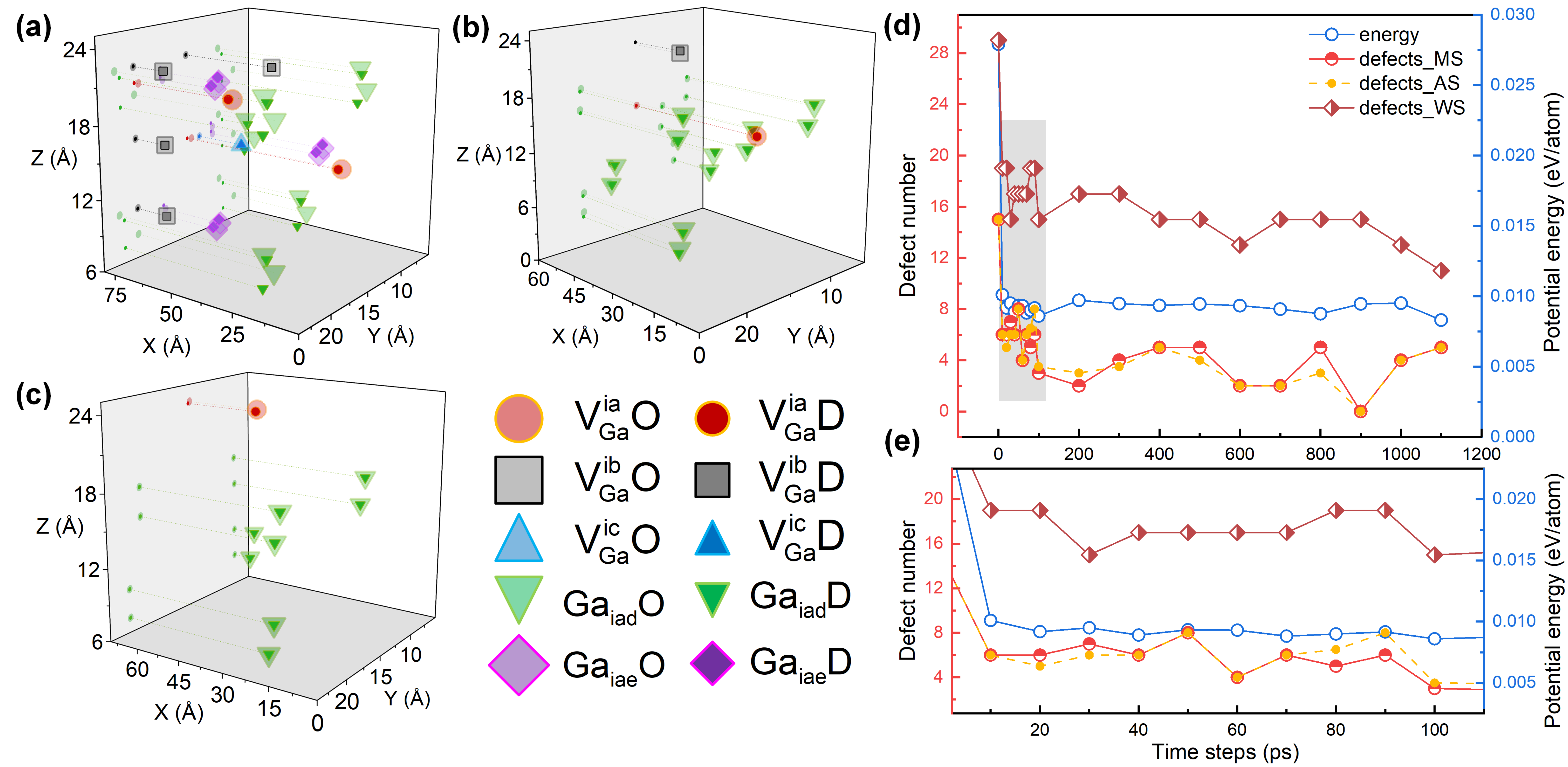}% Here is how to import EPS art
    \caption{\label{fig:anneal_defects}The evolutions of the potential energy and number of point defects during annealing at 900 K for 1.1 ns. 
    (a) The location and corresponding number of defects introduced at the beginning of setting up the test set. (b) The defect when annealed to 50 ps consists of 8 defects. (c) When annealed to 400 ps, the defect consists of 5 point defects.(d) The potential energies is calculated by further relaxing the corresponding frames to the local minimum at zero pressure and 0 K. `MS' represents the defect points counted by manual selection, and `AS' represents the defect points counted by the algorithm.`WS' represents the change in the number of all defects statistically obtained by the WS method. (e) Change in the number of defects from 1 to 100 ps. The initial 6001-atom cell consists of 7 Ga vacancy and 8 Ga interstitials.}
\end{figure*}
% 在采用了计算得到的放大阈值之后，通过实验，我们统计了对应不同缺陷构型的识别阈值的最小值，供参考，如表所示。对于Via构型，由于其原子环境和4配位的完美晶格Ga对应的环境十分相似，故算法十分容易将二者混淆，于是我们为ia构型设置的阈值为最大相似度的0.9倍，若小于该值便认为结构是其他构型。同理对于iad构型我们设置的识别阈值下限为最大识别结果的0.7倍；然而对于Vib，Vic和Iae构型经过大量实验验证，只要相似度大于0.5的类别基本上都能囊括所有正确的识别结果。最后在我们完成算法设计后，经过40组实验验证，发现对于离散的ia，ib和ic的点缺陷构型，该算法的识别准确率在95%即以上，对于ia而言为96%,ib为100%，ic构型为100%。这个准确率我们将随着实验的开展不断更新。
% 图5展示了不同的粒子总数，以及不同缺陷密度采用该算法得到的完美的识别结果，（a）图像粒子总数为4000，5中缺陷各一个得到的识别结果；（b）图像中粒子总数为4000，缺陷总数为10得到的识别结果；（c）图像中粒子总数为6000，缺陷总数为15和13得到的识别结果，可以发现对于离散且稳定的缺陷构型而言，该算法均能获得所有正确的结果。
% The program parameters include the expansion times of the lattice in different directions ($x, y, z$). 

% 而该过程研究的是低温条件下，稳定的晶格结构中氧化镓晶格对应的缺陷。实际上，为了推广算法的应用面，我们进一步探究的内容是，尝试将该方法运用至实验退火的过程中，也就是引入了不同的温度场，探究我们算法对实时缺陷的检测过程。并且从上述实验过程中，我们也发现了分裂间隙子构型与分裂空位构型有相互结合的趋势，组成的结构整体上呈现出无空位无间隙子的结构。而退火会扩大晶格修复的趋势，参照玻尔兹曼平衡方程，当空位结构与间隙子结构的数量近乎相同时，最后的构型应表现为无间隙子无空位结构。
% 首先，我们要探究温度场对键长的影响。
% 其次，统计得到函数关系后，引入膨胀因子γ，根据温度实时扩大数据库中的球壳体积，用以保证对应缺陷的特征曲线不发生变化。
% 在动态识别的实验结果，如图所示，其中缺陷点位随着退火时间的推移而逐渐减少，并且能量也随之逐渐降低，直到稳定结果。图中分别对应着在体系大小为4000时，间隙子总数为4，空位总数为4的实验一，间隙子总数为6，空位总数为6的实验二，另一种情况为体系大小为6000时，间隙子总数为10，空位总数为10的实验三，可以发现在实验过程中，体系均有缺陷构型与间隙子构型逐渐减少的趋势。
% 在我们的缺陷检测的动态平衡过程中，存在着空位缺陷与间隙子结构相互结合的情况，该过程形成了能量较低的稳定构型，但这种现象并不是该研究的重点于是我们并不在本文中进行该复杂结构的检测。并且我们实验发现，结合的空位结构与间隙子结构越多，该体系的能量就越低，可以从另一个角度来说明体系稳定和缺陷态的减少有关。对于结构所展示的光学与电学性质采用DFT计算方法和第一性原理结合的手段会更加高效。
During the dynamic equilibrium process of defect detection, 
a combination of vacancy defects and interstitials consistently arises, 
forming a stable configuration with low energy. 
However, as this phenomenon is not the primary focus of this study, 
we refrain from examining this complex structure in this paper. 
Additionally, tests in Fig.~\ref{fig:anneal_defects} reveal that the more vacancy and interstitial configurations are combined, 
the lower the energy of the system. 
This observation can be explained by considering the system's stability in relation to the reduction of defect states.

Lattice vibrations are particularly pronounced at high temperatures, 
leading to coordination number changes even in a perfect lattice. 
To mitigate the influence of lattice thermal vibrations, 
an energy minimization process is applied to the corresponding lattice information at a specific time step. 
This process yields a relatively stable structure used to test the algorithm's accuracy. 
Fig.~\ref{fig:anneal_defects} illustrates an annealing process in which 7 split vacancy defects and 8 split interstitials are introduced into another set of 6001 particles.

% 根据补充材料中所涉及的内容，我们发现在高温条件下，晶格振动十分剧烈，并且即使是完美晶格也存在部分点位的配位数的变化。于是为了减少晶格扰动对我们识别算法的影响，我们将特定时间步长下对应晶格信息进行能量最小化过程处理，可以得到较为稳定的结构，再使用该结构测试我们算法的准确性。统计得到如下的图像。

As depicted in Fig.~\ref{fig:anneal_defects}, 
over the 1.1 ns, 900 K annealing process, the system's structure gradually stabilizes, 
and the reduction in the number of defects, from 15 point defects to 5 point defects, 
reflects the stable state of system.
At the beginning of the annealing test, 2 $\mathrm{V}_\mathrm{Ga}^{ia}$ defects, 4 $\mathrm{V}_\mathrm{Ga}^{ib}$ defects, 1 $\mathrm{V}_\mathrm{Ga}^{ic}$ defect, 5 $\mathrm{Ga}_\mathrm{iad}$ defects, and 3 $\mathrm{Ga}_\mathrm{iae}$ defects are introduced into the system, totaling 6,001 particles, as shown in Fig.~\ref{fig:anneal_defects}a. 
In Fig.~\ref{fig:anneal_defects}b, after annealing at 50 ps, the number of point defects reduced to 8, comprising 1 $\mathrm{V}_\mathrm{Ga}^{ia}$ defect, 1 $\mathrm{V}_\mathrm{Ga}^{ib}$ defect, and 6 $\mathrm{Ga}_\mathrm{iad}$ defects. 
Following annealing at 400 ps, the number of point defects decreased to 5, including 1 $\mathrm{V}_\mathrm{Ga}^{ia}$ defect and 4 $\mathrm{Ga}_\mathrm{iad}$ defects, as shown in Fig.~\ref{fig:anneal_defects}c.

Throughout the annealing process, 
nearly all point defect configurations transform into $\mathrm{Ga}_\mathrm{iad}$ configurations and a composite configuration. 
Fig.~\ref{fig:anneal_defects}d illustrates the changes in the average energy of particles and the number of defects in the system after annealing at 1.1 ns, 900 K. The trends in energy variation and point defects closely align. 
The Fig.~\ref{fig:anneal_defects}d also presents the total number of defects calculated by the WS method in Open Visualization Tool (OVITO)~\cite{2010_ovito}. The total number of particles returned by this method is nearly twice that of the point defect configuration due to its diverse Voronoi space and the overestimation of interstitial and vacancy configurations.
In Fig.~\ref{fig:anneal_defects}e, the variation of the number of defects within 100 ps is partially magnified. 
Here, the total number of maunally calculated point defects closely matches the total number of point defects obtained by the algorithm. 
Manual select of defect changes and the results provided by our algorithm illustrate that the algorithm can accurately identify real-time results up to 88.2\%.
This observation indicates that our algorithm demonstrates excellent real-time performance in simplifying dynamic processes.

We note that the current research is limited to identifying intrinsic-defects, 
yet the exploration of material properties must also consider the impact of impurity atoms on optical and electrical properties. 
Subsequent work on the characteristics of doped $\beta$-\ce{Ga2O3} is anticipated to yield favorable results. 
Instead, it prioritizes the accuracy of the recognition results.

A challenge becomes apparent when the test sets contain both isolated point defect configurations and various densely packed point defect clusters. 
The substantial differences in similarity between defects may lead the algorithm to categorize them into two distinct groups, 
potentially excluding the lower similarity category from the final recognition results. 
This challenge is an area for future improvement, 
which could involve enriching the database information for each defect and employing more accurate environmental models for calculations.

\section{Conclusion}

In summary,
the ARDF function and similarity score designed by us, combined with particle swarm optimization algorithm and hierarchical clustering machine learning, achieve a very high accuracy of 95\% for Ga point defect configurations in the lattice of $\beta$-\ce{Ga2O3} in static processes.
For the complex structure formed by the combination of point defects, obtaining the defect configuration with certain position change is not achievable using our method.
Nevertheless, 
the randomly generating various intrinsic defects technique in large-scale $\beta$-\ce{Ga2O3} systems casts a new light on building extensive atomic database and
the combination of PSO and HC algorithms in our approach has opened avenues for future exploration to simulate crystal defect configurations. 
Our work offers a reliable method for identifying intricate defects in $\beta$-\ce{Ga2O3}.

\section*{Acknowledgments}

J. Zhao acknowledge the National Natural Science Foundation of China under Grant 62304097; Guangdong Basic and Applied Basic Research Foundation under Grant 2023A1515012048; Shenzhen Fundamental Research Program under Grant JCYJ20230807093609019. Project is also supported by State key laboratory of precision measuring technology and instruments (Tianjin University) under Grand pilab2203.
The study is supported by National Natural Science Foundation of China (No. 51575389, 51761135106), 2020 Mobility Programme of the Sino-German Center for Research Promotion (M-0396), and State key laboratory of precision measuring technology and instruments (Pilt1705, Pilt2107).

\bibliographystyle{apsrev4-2}
\bibliography{defect_detect}

\end{document}